%% file: main.tex
\documentclass[11pt]{article}

\usepackage[a4paper, total={6.5in, 8.5in}]{geometry}

\usepackage[english]{babel}

\usepackage[utf8]{inputenc}
\usepackage[english]{babel}
\usepackage[T1]{fontenc}

\usepackage{xcolor}
\usepackage{graphicx}
\usepackage{amsfonts, amsmath}
\usepackage{bbold}
\usepackage{framed}
\usepackage{listings}
\usepackage{import}

\usepackage{appendix}
\usepackage[colorlinks=true, allcolors=blue]{hyperref}
\usepackage{xspace}

\definecolor{shadecolor}{rgb}{0.94, 0.97, 1.0}
\providecommand{\pd}{\texttt{PulserDiff}\xspace}
\providecommand{\p}{\texttt{Pulser}\xspace}
\providecommand{\tor}{\texttt{PyTorch}\xspace}
\providecommand{\qm}{\texttt{QuantumModel}\xspace}
\providecommand{\bigT}{\boldsymbol{\Theta}}

\newcommand{\ket}[1]{| #1 \rangle}

\usepackage{tcolorbox}

\definecolor{bg}{gray}{0.95}

\title{\pd: a pulse differentiable extension for \p}

\author{
  Vytautas Abramavicius,
  Melvin Mathé,
  Gergana V. Velikova,
  João P. Moutinho,\\
  Mario Dagrada,
  Vincent E. Elfving,
  Alexandre Dauphin,
  Joseph Vovrosh,
  Roland Guichard
  \\
  Pasqal, 24 Av. Emile Baudot, 91120 Palaiseau, France
}
\date{}

\begin{document}
\maketitle

\begin{abstract}
Programming analog quantum processing units (QPUs), such as those produced by Pasqal, can be achieved using specialized low-level pulse libraries like \p. However, few currently offer the possibility to optimize pulse sequence parameters. In this paper, we introduce \pd, a user-friendly and open-source \p extension designed to optimize pulse sequences over a well-defined set of control parameters that drive the quantum computation. We demonstrate its usefulness through several case studies involving analog configurations that emulate digital gates and state preparation. \pd produces hardware-compatible pulses with remarkably high fidelities, showcasing its potential for advancing analog quantum computing applications.
\end{abstract}

\section{Introduction}

Pasqal's roadmap to manufacture quantum processing units (QPUs) leverages neutral-atom technology, where \textit{digital} and \textit{analog} operational modes can be implemented. Currently, accessible QPUs predominantly operate in the analog mode,  which governs the many-body dynamics of interacting rubidium (Rb) atoms confined within optical tweezers. It allows for the creation of highly configurable atomic layouts that underpin the system's computational capabilities \cite{henriet2020}. In principle, these atomic arrays can realize multiple different many-body Hamiltonians, most commonly considered of the Ising- or the $XY$ model-types. In this work, we adopt the Ising-type - hereafter referred to as the \textit{Rydberg Hamiltonian} - as our primary programming framework.

Programming these platforms requires specialized libraries that faithfully represent hardware configurations and degrees of freedom. One can cite Pasqal's low-level pulse programming library \p \cite{pulser2024} but also QuEra's \texttt{Bloqade} \cite{bloqade2023quera}, Q-CTRL's \texttt{Black Opal} \cite{blackopal} or \texttt{Qruise} \cite{qruise}. However, few offer the possibility to optimize built-in pulse control parameters, of utmost relevance to target neutral-atoms hardware amenable and noise-resilient pulse sequences.

This paper introduces \pd, a standalone extension to \p that allows for the optimization of a set of well-defined pulse control and register parameters. It is integrated with \tor \cite{Paszke2019} to leverage standard auto-differentiation (AD) functionality for machine-learning (ML) and optimization. As such, it extends \p types to \tor types, \textit{i.e.} \texttt{tensor} types for a straightforward integration with \tor functionalities and workflows. Moreover, \pd introduces a new \texttt{QuantumModel} type, which serves as a top-level entry point for applications to be derived from. 

The paper is structured as follows: in Section \ref{sec:ryd-ham} we recall the basics for the many-body dynamics of neutral-atom processors, namely the Rydberg Hamiltonian. In Section \ref{sec:implem} we present the software implementation of \pd. Section \ref{sec:res} shows results for research use cases on the emulation of digital gates and state preparation on analog platforms. Conclusions and future work are drawn in Section \ref{sec:conc}.
  
\section{The Rydberg Hamiltonian}
\label{sec:ryd-ham}

We focus here on the analog capabilities of neutral atom devices, which have garnered significant interest in both fundamental research in many-body physics \cite{browaeys2020, scholl2021}, and emerging industrial applications \cite{dalyac2025}. Notably, the Maximum Independent Set problem can be mapped directly onto neutral atom platforms \cite{ebadi2022, wurtz2024} (see \cite{pasqalmis} for an open-source implementation for Pasqal's hardware). These systems have also shown promise in various ML frameworks for quantum computations \cite{bravo2022, henry2021, leclerc2023, albrecht2023} and have been proposed for Variational Quantum Eigensolver (VQE) protocols that leverage the flexibility of lattice geometries \cite{michel2023, varia2024}. Furthermore, recent studies demonstrated their potential in drug discovery \cite{darcangelo2024}.
In neutral atom devices, qubits are encoded in two distinct electronic states of atoms, typically Rubidium or Strontium, that interact via a van der Waals potential. The states chosen to represent $\ket{0}$ and $\ket{1}$ can significantly affect interaction dynamics due to the atom's extensive range of electronic states. Here, we adopt a common encoding, where $\ket{0}$ is assigned to a low-energy state, and $\ket{1}$ to a Rydberg state. This configuration results in an Ising-type Hamiltonian governing the system’s interactions \cite{henriet2020} as

\begin{equation}
    \hat{H}(t) = \hbar \sum_j \frac{\Omega(t)}{2} \bigg[ 
\cos\big(\phi\big) \hat{\sigma}^x_j - 
\sin\big(\phi\big) \hat{\sigma}^y_j \bigg]
- \hbar \sum_j \frac{\delta_j(t)}{2} \hat{\sigma}^z_j
+ \sum_{i>j} \frac{C_6}{r_{ij}^6} \hat{n}_i \hat{n}_j,
\label{eq:neutral-atom-ham}
\end{equation}
where \( \hat{\sigma}^{x,y,z} \) are the Pauli \( x, y, z \) matrices acting on the \( i \)-th qubit, \( \hat{n}_i = \frac{1 + \hat{\sigma}_z}{2} \) is the number operator, \( r_{ij} \) is the distance between the \( i \)-th and \( j \)-th qubits, \( \phi \) is the phase, \( \Omega \) is the Rabi frequency, and \( \delta \) is the detuning of the external laser that couples the qubit ground and Rydberg states. These parameters can be interpreted as an effective magnetic field, with transverse and longitudinal components proportional to \( \Omega(t) \) and \( \delta(t) \), respectively. Each field can be tuned by adjusting the intensity and frequency of the laser field. The third term in Eq.~(\ref{eq:neutral-atom-ham}) describes interactions between individual atoms. Specifically, it accounts for the energy penalty incurred when two qubits are simultaneously in the Rydberg state, leading to the well-known Rydberg blockade~effect \cite{urban2009}. This coupling depends on the coefficient \( C_6 \), which is determined by the choice of Rydberg state \( n \).

\subsection{Optimizable parameters} \label{subsec:opt-params}

The Rydberg Hamiltonian in Eq.~(\ref{eq:neutral-atom-ham}) also represents a programmatic framework to drive the quantum computation on analog platforms through experimentally tunable and time-modulated laser control parameters $\Omega(t)$, $\delta(t)$, and $\phi$. Solving the Schr\"odinger equation for the Rydberg Hamiltonian involves time-propagating for a predefined time range $t\in[0,\tau]$ where the total pulse duration $\tau$ is also experimentally controllable. From a simulation point of view, it is therefore natural to investigate the development of a framework for the optimization of these pulse parameters as they straightforwardly mirror experimental ones. Moreover, one salient characteristic of neutral atom platforms is the high degree of register customization. This is represented through the customization of the interatomic distance $r_{ij}$ in Eq.~(\ref{eq:neutral-atom-ham}). Allowing for the optimization of $r_{ij}$ alongside the aforementioned pulse parameters enables the exploration of optimal register topologies prior to computations on (re)configurable platforms based on neutral atoms. This defines a relevant set of parameters of interest for a \pd implementation as
\begin{equation}
\Pi = \{ \Omega(t), \delta(t), \phi, r_{ij}, \tau\}.
\label{eq:params}
\end{equation}

\section{\pd implementation}
\label{sec:implem}

\subsection{\p recap} \label{subsec:pulser-recap}

Although released as an open-source standalone package, \pd \cite{pulserdiff} is tightly integrated and reliant on the \p Python package for register and pulse sequence creation. \p is designed for simulating and executing pulse sequences on neutral-atom quantum devices and consists of three main components: \texttt{pulser-core}, \texttt{pulser-simulation}, and \texttt{pulser-pasqal}. \texttt{pulser-core} is responsible for the implementation of low-level register and sequence creation and manipulation routines, but also device and noise specifications. \texttt{pulser-simulation} contains computational and numerical backends to handle the simulation of the constructed pulse sequence. \texttt{pulser-pasqal} provides capabilities to run pulse sequences through Pasqal's cloud services on either large-scale emulators or physical neutral-atom quantum devices. 

\p is tailored for analog quantum computations where users build a quantum program as a sequence of laser control pulses operating locally and/or globally on qubits contained in a register. Each pulse in the sequence is characterized by its amplitude $\Omega(t)$, detuning $\delta(t)$, and phase $\phi$, thus realizing a particular time-dependent instance of the Hamiltonian given in Eq.~(\ref{eq:neutral-atom-ham}). Several predefined pulse shapes are available in \p together with the possibility to arbitrarily customize pulses as well. Therefore, any $N$-pulse sequence represents a time-dependent piecewise Hamiltonian $\hat{H}_{\rm{seq}}(t)$ given by:

\begin{equation}
\hat{H}_{\rm{seq}}(t) = 
\begin{cases} 
\hat{H}_1(t), & t \in [0, t_1) \\
\hat{H}_2(t), & t \in [t_1, t_2) \\
\vdots & \vdots \\
\hat{H}_N(t), & t \in [t_{N-1}, \tau]
\end{cases}
\label{eq:seq-ham}
\end{equation}

This Hamiltonian is then sampled in \texttt{pulser-core} in time steps of 1 ns and passed to the selected computational backend of \texttt{pulser-simulation} that numerically solves either the Schr\"odinger or the Lindblad master equation (for noisy simulations) and outputs the final state (density matrix, respectively). \p supports sequence parameterization through its variable definition system. It allows for the creation of an abstract pulse sequence containing variable placeholders for pulse parameters, to be later filled with concrete values during the sequence-building process. Therefore, the same sequence template can be reused without recreation. All pulse shapes available in \p support parametrization for physical parameters of the Hamiltonian in Eq. (\ref{eq:neutral-atom-ham}) such as the amplitude $\Omega(t)$, detuning $\delta(t)$, phase $\phi$, and pulse duration $\tau$.

\subsection{From \p to \pd}
\label{subsec:from-pulser-to-pd}

\p's default numerical backend for performing quantum mechanical calculations relies on the \texttt{QuTiP} package \cite{qutip2012}, itself built on top of the computationally efficient package \texttt{NumPy} \cite{numpy2020}. Although \texttt{NumPy} supports finite-difference numerical differentiation, it does not support AD methods, favored in modern ML processes due to their robustness against errors. Consequently, the \texttt{QuTiP} backend cannot serve as an implementation that supports AD with respect to pulse and register parameters defined in $\Pi$ [see Eq.~\eqref{eq:params}]. To achieve this, \pd implements a \tor-based backend that seamlessly integrates with \texttt{pulser-core},  allowing for the creation of a computational graph where user-defined pulse and register parameters are leaf nodes. Subsequently, \tor's \texttt{autograd} engine uses the chain rule to calculate the required gradients which are finally passed to any compatible gradient-based optimizer, \textit{e.g.} \texttt{Adam} \cite{kingma2017adam}. 

\subsection{The \texttt{QuantumModel}}
\label{subsec:qm}

\pd provides a central user-facing \texttt{QuantumModel} type to enhance \p's pulse sequence and register objects with \tor features. It subclasses \tor's standard \texttt{nn.Module} for complete compatibility with other components (optimizers, schedulers, and loss functions) to build standard ML pipelines. 

The abbreviated signature of the \qm class is presented in Code sample~\ref{code:QM-sign}. \qm initialization relies mainly on the predefined \p sequence object \texttt{seq} for which optimizable (trainable) parameters names and initial values are provided in the \texttt{trainable\_param\_values} dict, together with custom waveform functions (see Section \ref{sec:custom-wf}). Optionally, a \texttt{constraints} dict can be passed to account for min/max hardware specification ranges.

\begin{shaded*}
\begin{lstlisting}[caption=Abbreviated signature for the \qm class initialization., label={code:QM-sign}, language=python]
class QuantumModel(nn.Module):
    def __init__(
        self,
        seq: Sequence,
        trainable_param_values: dict[str, Tensor] | dict[str, tuple[tuple, Callable]] = {},
        constraints: dict[str, Any] = {},
        ...
    ) -> None:
\end{lstlisting}
\end{shaded*}

Behind the scenes, \qm arguments are registered as trainable \texttt{nn.Module} attributes, exposing them for subsequent update inside an optimization loop. Moreover, trainable parameters are matched with variables defined in the abstract \texttt{seq} object, for its final concretization with initial parameter values. The resulting sequence is then simulated using the \texttt{forward()} method, which outputs a list of state vectors or density matrices evaluated at each discrete time step $t_i$. The final state evaluated at $\tau$ is then used in a standard loss calculation and optimization pipeline.

\subsection{Optimizing parameters} \label{subsec:opt-loop}

To illustrate \pd optimization capabilities, we rely on solving a toy problem: for a given pulse sequence and input state, find a set  of pulse and/or register parameters $\bigT$ such that the expectation value $f(\bigT)=\left\langle \psi(\bigT)\right|\hat{C}\left|\psi(\bigT)\right\rangle $ for some observable $\hat{C}$ approaches a target value $f_{\rm{target}}$. The general optimization pipeline consists of four steps:

\begin{enumerate}
    \item Define a \p register and sequence representing the system to optimize. This part follows the standard \p syntax and uses its built-in constructs for pulses and sequence creation. 
    \item Create a \texttt{trainable\_param\_values} dictionary of pairs of parameters to optimize and initial values as \tor tensors. Define any additional constraints on the range of these parameters in \texttt{constraints}.
    \item Create a \qm instance by passing the pulse sequence and a parameter dictionary with optional constraints. This part bridges \p and \tor frameworks.
    \item Define a \tor optimizer and execute an optimization loop using the \qm. This part follows standard \tor practices for ML model training/optimization.
\end{enumerate}

In following, we outline the flexibility of the optimization process for chosen pulse parameters using both predefined and custom waveforms. We also broach the optimization of the total pulse duration, currently only enabled for constant pulse waveforms. Finally, we address the problem of register parameters optimization.

\subsubsection{Pulse parameter optimization for predefined waveforms} \label{sec:pulse-param-opt}

A standard \p workflow for the definition of a two-qubit register and a sequence containing two pulses is presented in Code sample \ref{code:reg-seq-creation}. The key element is the declaration of \p variables as placeholders for optimizable parameters to be later filled with concrete values.

\begin{shaded*}
\begin{lstlisting}[caption=Register and pulse sequence creation using \p., label={code:reg-seq-creation}, language=python]
# Create a pulser register with a rectangular layout.
reg = Register.rectangle(1, 2, spacing=8, prefix="q")

# Create a pulser sequence and declare channels.
seq = Sequence(reg, MockDevice)
seq.declare_channel("rydberg_global", "rydberg_global")

# Declare sequence variables for pulse parametrization.
omega_param = seq.declare_variable("omega")
area_param = seq.declare_variable("area")

# Create pulser parametrized pulses with pre-defined waveforms.
pulse_const = Pulse.ConstantPulse(1000, omega_param, 0.0, 0.0)
amp_wf = BlackmanWaveform(800, area_param)
det_wf = RampWaveform(800, 5.0, 0.0)
pulse_td = Pulse(amp_wf, det_wf, 0)

# Add pulses to the sequence with appropriate channels.
seq.add(pulse_const, "rydberg_global")
seq.add(pulse_td, "rydberg_global")
\end{lstlisting}
\end{shaded*}

The second step is the instantiation of the \qm class with relevant arguments (see Code sample \ref{code:qm-instance}). It results in a \tor \texttt{nn.Model} object and associates declared \p variables (presently \texttt{omega\_param} and \texttt{area\_param}) with respective \tor tensor values passed in the \texttt{trainable\_params} dict.  

\begin{shaded*}
\begin{lstlisting}[caption=Creation of \qm instance., label={code:qm-instance}, language=python]
# Define pulse parameters as torch.Tensor
omega = torch.tensor([5.0], requires_grad=True)
area = torch.tensor([torch.pi], requires_grad=True)

# Create trainable parameters and constraints dicts.
# Names must match the pulser variables.
trainable_params = {"omega": omega, "area": area}
constraints = {
    "omega": {"min": 4.5, "max": 5.5}
}

# Create quantum model from sequence, parameter values, and constraints.
model = QuantumModel(seq, trainable_params, constraints)
\end{lstlisting}
\end{shaded*}

\begin{figure}[h]
\centering
\includegraphics[scale=0.75]{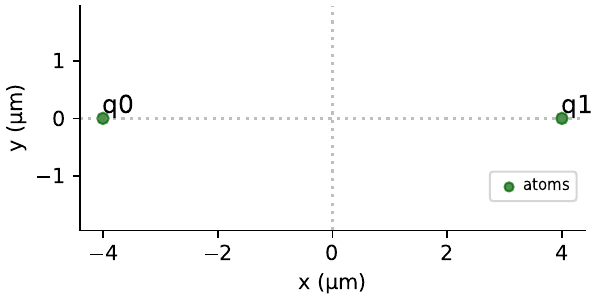}
\caption{2-qubit register.}
\label{fig:2q-reg}
\end{figure}

\begin{figure}[h]
\centering
\includegraphics[width=\textwidth]{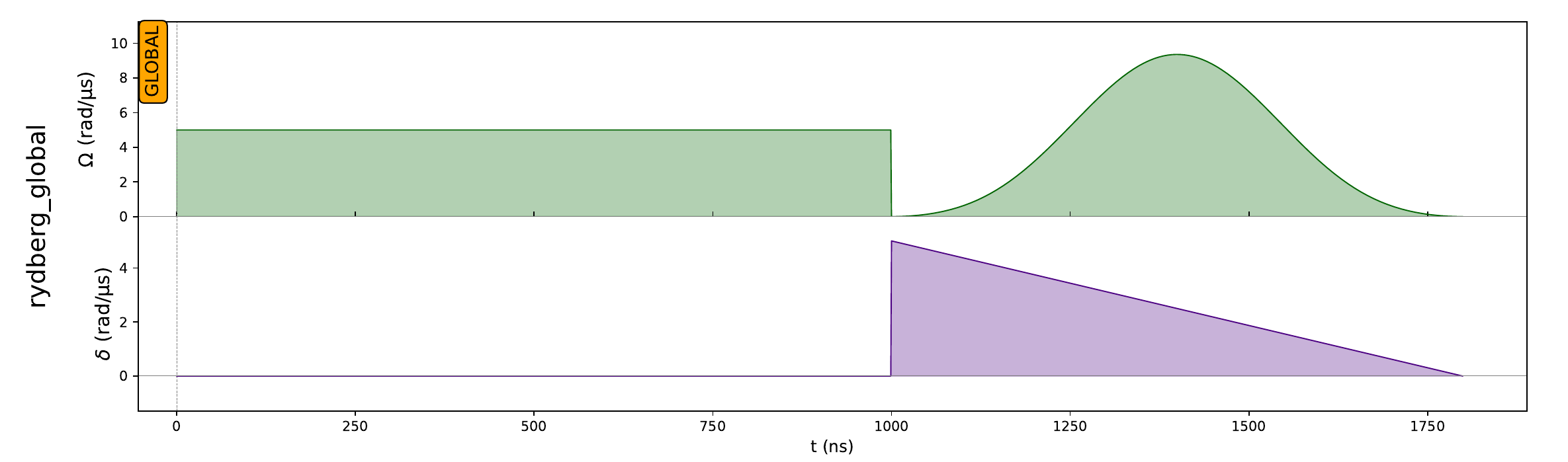}
\caption{Initial sequence containing one constant and one time-dependent pulse.}
\label{fig:2q-seq}
\end{figure}

The register and initial sequence visualizations are shown in Figs. \ref{fig:2q-reg} and \ref{fig:2q-seq}, respectively. Finally, we can construct and execute an optimization loop as described in Code sample \ref{code:opt-loop}. It follows a standard ML training pipeline together with few \pd specificities. For instance, the \texttt{model.check\_constraints()} call restricts the range of optimizable parameters solutions according to values provided in the \texttt{constraints} dict. The \texttt{model.update\_sequence()} call rebuilds the underlying parametrized \p sequence with iteratively updated parameter values. 

\newpage

\begin{shaded*}
\begin{lstlisting}[caption=Optimization loop for the \qm., label={code:opt-loop}, language=python]
# Define a loss function and an optimizer from torch
# and a target value.
loss_fn = torch.nn.MSELoss()
optimizer = torch.optim.Adam(model.parameters(), lr=0.05)
target_value = 1.e-6

# Optimize model parameters until the final output expectation value
# matches the target value.
for t in range(epochs):
    # Compute expectation and loss from the final state vector.
    _, exp_val = model.expectation()
    loss = loss_fn(exp_val.real[-1], target_value)

    # Standard backpropagation.
    loss.backward()
    optimizer.step()
    optimizer.zero_grad()

    # Enforce constraints on optimizable parameters.
    model.check_constraints()

    # Update sequence with optimized pulse parameter values.
    model.update_sequence()
\end{lstlisting}
\end{shaded*}

After the optimization loop has completed, a generator for optimal parameters can be extracted using \texttt{model.named\_parameters()} or the final sequence can be accessed through the \texttt{model.built\_seq} attribute.

\subsubsection{Custom waveform optimization} \label{sec:custom-wf}

In situations where finer-grained control over pulse shapes is required, parametrized pulses with predefined waveforms is not sufficient. Still, \p provides a \texttt{CustomWaveform} type where an arbitrary array of values is interpolated to reconstruct the waveform. The process remains identical to the previously stated workflow: declare a variable associated with the custom waveform and create the corresponding pulse (see Code sample \ref{code:opt-wf}). 

\begin{shaded*}
\begin{lstlisting}[caption=Creation of pulse with optimizable waveform in \p., label={code:opt-wf}, language=python]
# Declare parameters for the amplitude and detuning waveforms. 
pulse_duration = 300  # ns
custom_omega = seq.declare_variable("custom_omega", size=pulse_duration)

# Create a custom-shaped amplitude and constant detuning waveforms.
custom_amplitude_wf = CustomWaveform(custom_omega)
constant_detuning_wf = ConstantWaveform(pulse_duration, 1.5)

# Create a custom-shaped pulse from both waveforms above.
custom_pulse = Pulse(custom_amplitude_wf, constant_detuning_wf, 0.0)
\end{lstlisting}
\end{shaded*}

The main difference from the usual \p variable declaration is the \texttt{size} argument. This way, \p is instructed to reserve an array of values of length \texttt{pulse\_duration} for the custom amplitude waveform.  In \pd, this custom waveform is used in a very similar manner: the waveform is defined by creating a parametrized function $w_{\theta}(t)$ in the interval $[0, \tau]$ where $\tau$ is the total duration of the pulse (see Code sample \ref{code:custom-wf-func}).

\begin{shaded*}
\begin{lstlisting}[caption=Function definition for the custom waveform in \pd., label={code:custom-wf-func}, language=python]
# Define trainable pulse parameters as torch.Tensor.
param1 = torch.tensor(6.0, requires_grad=True)
param2 = torch.tensor(2.0, requires_grad=True)

# Define a custom amplitude waveform.
def custom_wf(param1, param2) -> torch.Tensor:
    x = torch.arange(pulse_duration) / pulse_duration
    return param1 * torch.sin(torch.pi * x) * torch.exp(-param2 * x)

# Add optimizable pulse parameters.
trainable_params = {
    "custom_omega": ((param1, param2), custom_wf),
    ...
}
\end{lstlisting}
\end{shaded*}

 In this example, \texttt{param1} and \texttt{param2} arguments are used to control the shape of the underlying waveform returned as a \tor tensor of the same length as the previously defined \p variable. This custom waveform function can now be added as another entry to the \texttt{trainable\_params} dict and passed to the \qm constructor. In this case, optimizable parameters are control arguments for the waveform function. The actual waveform tensor is computed by the \qm instance at runtime. The syntax for passing optimizable parameters is slightly different since one must provide a tuple of parameters together with the required waveform function for evaluation. 

\subsubsection{Total duration optimization}

As usual, \pd supports pulse duration optimization through \p duration parametrization. However, the current version only enables it for sequences composed of piecewise constant pulses. From a user perspective, declaring an optimizable duration parameter follows a similar syntax as for any other declarable \p parameter as shown in Code sample \ref{code:opt-dur}.\\

\begin{shaded*}
\begin{lstlisting}[caption=Declaration of optimizable pulse duration parameter., label={code:opt-dur}, language=python]
# Declare a pulse duration variable.
duration = seq.declare_variable("duration")

# Create a duration parameterized pulse.
...

# Define pulse duration parameter as torch.Tensor.
# /!\ duration unit is us not ns.
duration = torch.tensor([0.4], requires_grad=True)

# Add optimizable pulse and register parameters.
trainable_params = {
    "duration": duration,
    ...
}
\end{lstlisting}
\end{shaded*}

One important caveat here is the unit for the initial duration value in $\mu$s when standard \p accepts integer values for durations in ns. This particular choice of units circumvents the fact that the \tor AD engine does not support tensors of integer data type for gradient computations. Another caveat is related to the inner sampling mechanism used in \texttt{pulser-core} to discretize the pulse sequence. During this process, the sequence is sampled in fixed 1 ns time steps, thus constructing a discretized representation of the Rydberg Hamiltonian as in Eq.~(\ref{eq:seq-ham}). However, the \tor AD engine optimizes and updates continuous parameters in arbitrary increments and cannot account for the pulse duration parameter with fixed increments. \pd provides a solution by transforming the input \p sequence of constant pulses into an auxiliary sequence composed of 1 ns constant pulses with overall amplitude, detuning, and phase envelopes reproducing the original sequence. These envelopes are smooth parametrized functions that include duration parameters as floating point-type for \tor tensors to be optimized. 

\subsubsection{Register parameter optimization}

In the previous sections, we presented a general outline to solve an optimization problem for pulse parameters. However, neutral-atom technology offers to arbitrarily customize the register geometry, of utmost relevance for the execution of algorithms based on graph-structured data. In reflection, \pd offers the possibility to optimize the register topology through the optimization of interatomic distances. To illustrate this, let us revisit the toy problem with explicit qubit coordinates registered as new optimizable parameters \texttt{"q0"} and \texttt{"q1"} (see Code sample \ref{code:opt-reg-creation}). Note, that for qubit coordinates to be optimizable, the corresponding values must be passed explicitly to the \texttt{Register} constructor. 

\begin{shaded*}
\begin{lstlisting}[caption=Optimizable register creation with explicit atomic coordinates., label={code:opt-reg-creation}, language=python]
# Create a register with explicit and trainable tensor coordinates.
q0_coords = torch.tensor([0.5, 0.4], requires_grad=True)
q1_coords = torch.tensor([8.3, 0.1], requires_grad=True)
reg = Register({"q0": q0_coords, "q1": q1_coords})

# Adding optimizable pulse and register parameters.
trainable_params = {
    "omega": omega,
    "area": area,
    "q0": q0_coords,
    "q1": q1_coords,
}
\end{lstlisting}
\end{shaded*}

The further optimization procedure remains identical to the one described in the Code sample \ref{code:opt-loop}.


\section{Results}
\label{sec:res}

To illustrate \pd efficiency and relevance, we apply it for research use-cases in analog quantum computation without loss of generality. The first use case aims at unitary optimization: authors in \cite{varia2024} devise a scheme to emulate single qubit digital gates by optimizing global rotations together with single qubit gates in an analog (always-on) interaction framework (see Subsection \ref{subsec:global-optim}). In this case, the better the optimization quality for the global rotations, the better the fidelity in the digital emulation. The simulated unitary is reconstructed at the final step of the differentiable time-evolution for all possible combinations of input states. This requires some computational power and is currently restricted for a handful of qubits within a register for runnable simulations on standard laptops. However, \pd is designed to scale on High-Performance Computation environments that offer larger computational resources. A second use case aims at state preparation: the optimization of global rotations is performed to realize some ideal target state (see Subsection \ref{subsec:state-prep}). It is a computationally less costly optimization case as it only requires simulations for a selected single input state of interest.

\subsection{Global rotations optimization}
\label{subsec:global-optim}

We present here a general use case for the emulation of digital quantum gates in an analog context \cite{varia2024}. The main idea is to optimize parametrized pulses that drive the time evolution of the given interacting Rydberg Hamiltonian using \pd functionalities for a global (Hadamard) $\frac{\pi}{2}$ rotation as target unitary. The optimization process starts from an ansatz pulse, computes the fidelity of the simulated unitary by comparing with the target unitary. Then, a gradient descent-based algorithm can be executed until convergence is achieved. Two exploratory avenues were undertaken: first for hardware-constrained piecewise-constant waveforms, or square pulses, as a proof-of-principle for the retrieval of results in \cite{varia2024}, and second, for hardware-constrained custom waveforms. 

\subsubsection{Global rotation optimization with square pulses} \label{sec:gate-opt-const}

As previously mentioned, the emulation of digital gates was shown to be possible with piecewise-constant pulses with fidelities above 99\% for a linear layout of qubits with nearest-neighbour interactions \cite{varia2024}. In this work, we use \pd optimization capabilities to solve a similar problem, with some noticeable differences:

\begin{itemize}
    \item A \tor-based optimizer \texttt{Adam} is used instead of the \texttt{L-BFGS-B}.
    \item The targeted unitary is a global Hadamard gate instead of an \texttt{RX} gate to prepare a maximally superposed state from the initial state.
    \item Qubit interaction is not limited to nearest-neighbours.
\end{itemize}

Following \cite{varia2024} and the workflow in Section \ref{sec:pulse-param-opt}, our aim is to optimize parameters for a predefined \p sequence containing $K=8$ constant pulses of equal duration $\tau_{\rm{pulse}}=131$ ns and a $N=2$ qubits register arranged in a linear layout. For this simulation, the interatomic distance is set to $r_{12}=6.5\,\mu$m and each pulse is parametrized by its amplitude $\Omega_i$, detuning $\delta_i$ and phase $\phi_i$. The optimizable pulse sequence is then constructed in the \qm instance together with provided initial parameters values $\{\Omega_i, \delta_i, \phi_i\}_{i=1}^N$ and corresponding hardware constraints. The constraints for the amplitude are $\Omega_i \in [0, 4\pi]$ $\rm{rad}\cdot\mu\rm{s}^{-1}$ and the detuning $\delta_i \in [-4\pi, 4\pi]$ $\rm{rad}\cdot\mu\rm{s}^{-1}$. The simulated unitary $U_{\rm{sim}}$ is reconstructed at each optimization step for all possible $N$-qubits computational states passed as a batched initial state to the \qm. $U_{\rm{sim}}$ is compared to the ideal global Hadamard unitary $U_{\rm{target}}$ using the fidelity $F(U_{\rm{sim}}, U_{\rm{target}})=\frac{1}{2^N}\left|Tr(U_{\rm{sim}}^\dagger U_{\rm{target}})\right|$ which defines the loss function $L=1-F$ inside the optimization loop. Details of the sequence construction and optimization can be found in Appendix \ref{appendix:appA}. Fig. \ref{fig:2q-gate-opt} displays the resulting optimized sequence for a fidelity of $99.54\%$.

\begin{figure}[h]
\centering
\includegraphics[width=\textwidth]{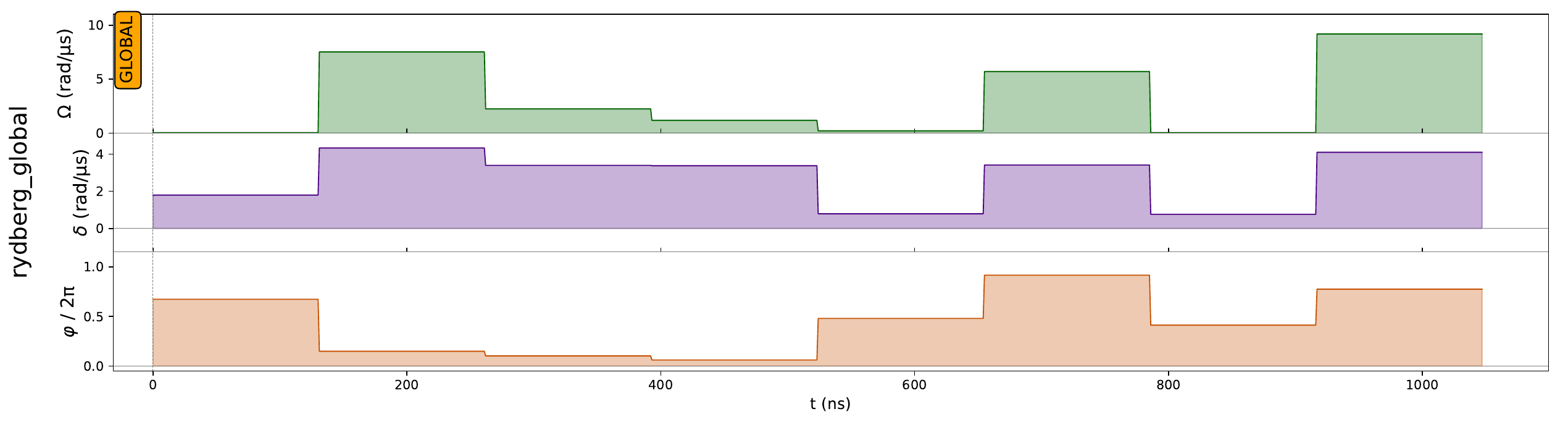}
\caption{The optimized sequence for 8 constant pulses and a 2-qubit linear register layout.}
\label{fig:2q-gate-opt}
\end{figure}

Table \ref{tab:square-pulses-fid} reports fidelities for varying register size. The corresponding optimized sequences are presented in Appendix \ref{appendix:appB}. The general fidelity decreasing trend can be attributed to long range interactions present in the Rydberg Hamiltonian that increase qubit couplings while scaling the system size. It incurs an increased complexity in the loss function landscape that the optimizer has to navigate resulting in lower quality solutions. More sophisticated optimization loops might help mitigating such effect by enabling optimizable degrees of freedom and relaxing constraints such as pulse waveforms and durations, as demonstrated in the next section.

\begin{table}[h]
\centering
\begin{tabular}{|c|c|}
\hline
\textbf{\# of qubits} & \textbf{fidelity, \%}  \\
\hline
2 & 99.54  \\
3 & 98.09  \\
4 & 97.31  \\
5 & 94.08  \\
6 & 93.11  \\
7 & 95.25  \\
\hline
\end{tabular}
\caption{Fidelities for optimized constant-pulse unitaries $U_{\rm{sim}}$ and different linear register sizes with .}
\label{tab:square-pulses-fid}
\end{table}

\subsubsection{Global rotation optimization with continuous pulses} \label{sec:gate-opt-cont}

Ideal constant pulses are not faithfully implementable on a real quantum device mainly due to sharp rises and falls. Therefore, the optimization problem should provide more realistic, smoother pulse shapes that are hardware amenable. To achieve this, we define first the total duration of the pulse $\tau$ and $M$ control parameters $\theta_m$ to customize arbitrary pulse waveforms $w(t)$ for the amplitude $\Omega(t)$ or the detuning $\delta(t)$. They are evaluated at uniformly spaced grid time points $t_m$ with time step $\Delta=\frac{\tau}{M+1}$ such that $w(t_m)=\theta_m$ together with boundary conditions $w(0) = w(\tau) = 0$. Then, the whole waveform is reconstructed by sine-interpolating these sampled points. 

Let us focus on the interpolation procedure. The goal is to calculate the value $w(t)$ given an arbitrary time instance $t$ using sampled points $\{\theta_m\}$ and ensure continuity and smoothness of the resulting function. First, we can calculate the interval index $m = \lfloor \frac{t}{\Delta} \rfloor$ such that $t\in [t_m,t_{m+1}]$. In this case, $w(t)$ values can be interpolated using the sine transition function:    

\begin{equation}
s(h) = \frac{1 + \sin(\pi h - \frac{\pi}{2})}{2}.
\label{eq:sine-fun-interp}
\end{equation}

\noindent This function transitions smoothly from 0 to 1 as $h \in [0, 1]$, and has continuous first derivatives. For $t\in [t_m,t_{m+1}]$, $w(t)$ values are calculated as:

\begin{equation}
w(t) = \theta_m (1 - s(h_{t,m})) + \theta_{m+1} s(h_{t,m})
\label{eq:interp-formula}
\end{equation}

\noindent where $h_{t,m} = \frac{t - t_m}{\Delta} \in [0, 1]$. This ensures that $w(t)$ is continuously differentiable and satisfies the interpolation condition $w(t_m) = \theta_m$. We can see from Eq. (\ref{eq:interp-formula}) that for any time $t$ the value $w(t)$ is a linear combination of only two values from the set of the control parameters $\{\theta_m\}$. Furthermore, as explained in Section \ref{subsec:pulser-recap}, any pulse sequence in \pd is discretized internally into 1 ns steps, thus we need to perform interpolation only for a discrete set of times $t_k=k$ with $k \in \{0, 1, ..., \tau-1\}$. Discreteness and linearity of Eq. (\ref{eq:interp-formula}) suggest that it can be rewritten  as a matrix-vector product. To achieve this we define a matrix $\mathbf{A} \in \mathbb{R}^{\tau \times M}$ such that each row contains only two non-zero entries $A_{k,m}=(1 - s(h_{k,m}))$ and $A_{k,m+1}=s(h_{k,m+1})$. After arranging the control parameters $\theta_m$ into a column vector, we obtain the following expression for the value of the waveform $w(t_k)$:

\begin{equation}
w(t_k) = A_{k,m} \theta_m + A_{k,m+1}\theta_{m+1} = (\mathbf{A} \cdot \mathbf{\theta})_k.
\label{eq:interp-formula-mat}
\end{equation}

Closer examination of the interpolated waveform in Eq. (\ref{eq:interp-formula-mat}) reveals that it guarantees $w(t_k)$ values to lie in the interval $[\rm{min}(\{\theta_m\}), \rm{max}(\{\theta_m\})]$, thus being bounded by control parameters. This feature can be exploited to implement external constraints on the pulse shape, currently not supported out-of-the-box by \pd custom waveforms. For standard hardware constraints, the amplitude waveform must be bounded in the interval $[0, \Omega_{\rm{max}}]$ and the detuning waveform in the interval $[-|\delta_{\rm{max}}|, |\delta_{\rm{max}}|]$. To achieve this, $\theta_m$ values are transformed with some bounded continuous and differentiable function to be then plugged in Eq. (\ref{eq:interp-formula-mat}). 

Here, we chose the sigmoid function $\sigma(x)$ to bound the amplitude. The waveform control parameters $\theta_{m,\Omega}$ then transform into:

\begin{equation}
\theta_{m,\Omega}^{\prime} = \Omega_{\rm{max}}\sigma(\gamma_\Omega\theta_{m,\Omega}).
\label{eq:transf-func-amp}
\end{equation}
The factor $\gamma_\Omega$ (usually smaller than 1) controls the slope of the sigmoid and mitigates the vanishing gradient problem during optimization. 
The corresponding transformation for the detuning uses the hyperbolic tangent $\rm{tanh}$ for detuning values to be negative. The resulting expression for the control values $\theta_{m,\delta}$ of the detuning waveform is then 

\begin{equation}
\theta_{m,\delta}^{\prime} = |\delta_{\rm{max}}|\rm{tanh}(\gamma_\delta\theta_{m,\delta}).
\label{eq:transf-func-det}
\end{equation}

Implementation details of the custom pulse optimization problem are given in Appendix \ref{appendix:appA}. Fig. \ref{fig:4q-gate-opt} shows the final optimized pulse for a system with a 4-qubit linear register, $\tau=1100$ ns, $M=20$, $\Omega_{\rm{max}}$, $|\delta_{\rm{max}}|$ are both set to $4\pi$ $\rm{rad}\cdot\mu\rm{s}^{-1}$ and the transformation parameter $\gamma$ from Eqs. (\ref{eq:transf-func-amp}-\ref{eq:transf-func-det}) is set to 0.05 for both amplitude and detuning waveforms. We can notice that the pulse shapes exhibit a  symmetry similar to the one found in \cite{varia2024}.

\begin{figure}
\centering
\includegraphics[width=\textwidth]{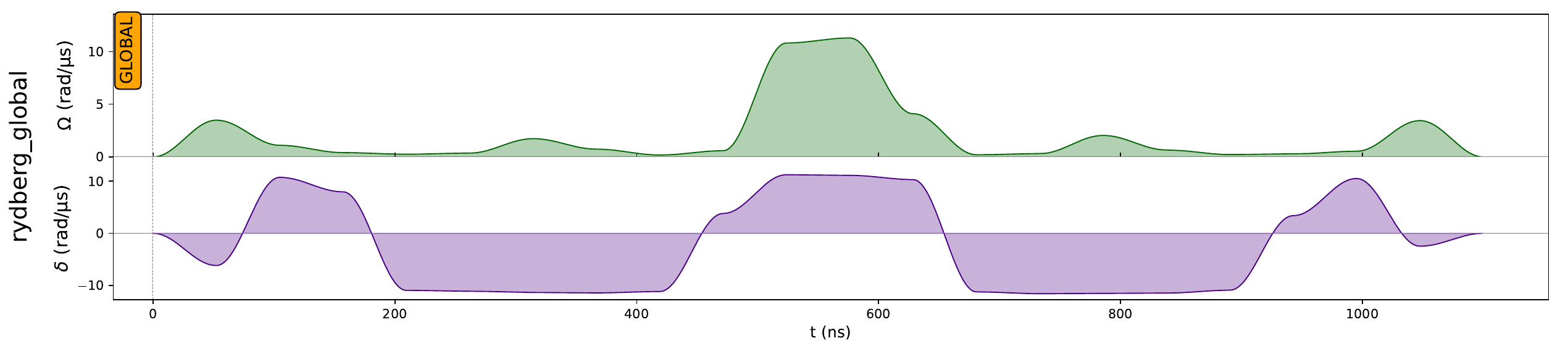}
\caption{Visualization of the optimized sequence consisting of the custom-waveform pulse for the gate optimization problem with a linear 4-qubit register.}
\label{fig:4q-gate-opt}
\end{figure}

Pulse shape optimization simulations were also performed for systems with different numbers of qubits in the register as reported in Table~\ref{tab:custom-pulses-fid}. The corresponding optimized sequences are presented in Appendix \ref{appendix:appB}. The obtained fidelities are all remarkably above $99\%$ showcasing an 5-6\% increase compared to fidelities calculated using constant pulses (see Section \ref{sec:gate-opt-const}) for systems with $N\geq5$ qubits. We also can notice that the fidelity decreases much slower with custom pulses, possibly due to increased flexibility offered by the custom pulse shape to mitigate the effects of the long-range inter-qubit interaction. This fidelity degradation is also reported in \cite{varia2024} to worsen by 2-3\% for larger systems.

\begin{table}
\centering
\begin{tabular}{|c|c|}
\hline
\textbf{\# of qubits} & \textbf{fidelity, \%}  \\
\hline
2 & 99.99  \\
3 & 99.84  \\
4 & 99.85  \\
5 & 99.73  \\
6 & 99.61  \\
7 & 99.51  \\
\hline
\end{tabular}
\caption{Fidelities of the optimized custom-pulse unitaries $U_{\rm{sim}}$ and different linear register sizes.}
\label{tab:custom-pulses-fid}
\end{table}


\subsection{State preparation}
\label{subsec:state-prep}

A vast majority of quantum algorithms require preparing specific initial states prior to execution for optimal computation. In this section, we show how an arbitrary state can be prepared using \pd sequence optimization capabilities. The optimization setup is indeed very similar to the one used in Section \ref{sec:gate-opt-cont}: the sequence consists of a single smooth and continuous I custom pulse with optimizable amplitude and detuning waveforms (see Appendix \ref{appendix:appA} for details). In this simulation, we choose the state $\left|1\dots1\right\rangle $ as the target state to prepare, whereas we initialize the optimization process with the zero state $\left|0\dots0\right\rangle $. A metric for state similarity is defined as the state fidelity $F=|\left.\left\langle \psi_{\rm{final}}\right| \psi_{\rm{target}} \right\rangle |^2$ which serves to define the optimization loss as $L=1-F$. 

\subsubsection{State preparation for linear layout}
\label{subsubsec:state-prep-linear}

\begin{figure}
\centering
\includegraphics[width=\textwidth]{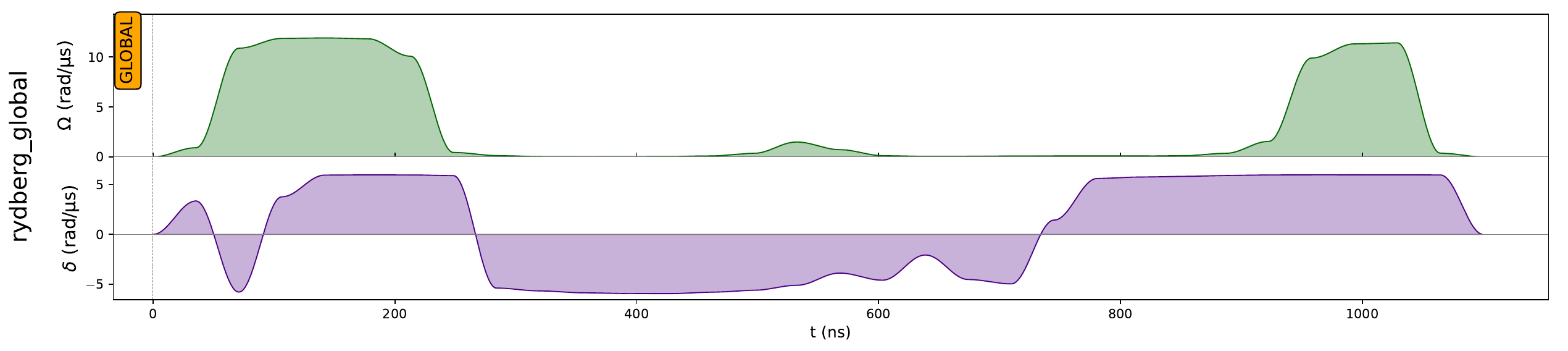}
\caption{Visualization of the optimized sequence consisting of custom-waveform pulse for the state preparation problem with a linear 6-qubit register.}
\label{fig:6q-state-prep}
\end{figure}

The optimized sequence for a linear register of 6 qubits is given in Fig. \ref{fig:6q-state-prep}. The chosen Rydberg level is $n=60$, the inter-atomic distance of the linear register is set to 7 $\rm{\mu m}$ and the pulse length is 1100 ns. We also set the hardware constraints $\Omega_{\rm{max}}=4\pi$ $\rm{rad}\cdot\mu\rm{s}^{-1}$ and $|\delta_{\rm{max}}|=2\pi$ $\rm{rad}\cdot\mu\rm{s}^{-1}$. The number of control parameters for the amplitude and detuning waveforms is set to $M=30$. With given register parameters, the ratio of nearest-neighbor interaction energy $J$ to the maximal amplitude $\Omega_{\rm{max}}$ reached by the optimized pulse is $\frac{\Omega_{\rm{max}}}{J} \approx 0.2$ which indicates that the interaction term in Eq. (\ref{eq:neutral-atom-ham}) cannot be neglected since it greatly influences the dynamics of the system. However, optimization of the pulse shape performed with \pd allowed to achieve the final fidelity value of $99.85\%$ which demonstrates its efficiency in compensating for the interaction effects. 

To further demonstrate the possibilities of the state preparation facilitated by \pd we performed similar simulations for different number of qubits using the linear register layout. The results are given in Table \ref{tab:state-prep-fid-linear} and the corresponding pulse shapes are shown in Appendix \ref{appendix:appC}.

\begin{table}
\centering
\begin{tabular}{|c|c|}
\hline
\textbf{\# of qubits} & \textbf{fidelity, \%}  \\
\hline
2 &  $>99.99$  \\
3 &  98.87  \\
4 &  96.09  \\
5 &  93.55  \\
6 &  99.85  \\
7 &  99.76  \\
\hline
\end{tabular}
\caption{Fidelities of the state $\left|\psi_{\rm{final}}\right\rangle $ after the pulse optimization with linear register layout.}
\label{tab:state-prep-fid-linear}
\end{table}

Table \ref{tab:state-prep-fid-linear} reveals that fidelity decreases from 99.99 \% for 2-qubit linear register to 93.55 \% for a 5-qubit register, however, regains values above 99 \% for 6 and 7 qubits. The obtained fidelity values are quite respectable, however, the reason for the observed decrease is not immediately clear but might be related to a complex interplay between a  particular set of parameters for the system size and the pulse duration. Such investigation of the influence of the pulse duration and register layout on the state fidelity is briefly presented in the next section.  

\subsubsection{State preparation for custom layout and different pulse durations}

In Table \ref{tab:state-prep-fid-diff-durations} we present the fidelity values simulated with shorter (1000 ns) and longer (1200 ns) custom pulses than in the previous section. We can see that with the exception of the 2-qubit system, for all other register sizes the fidelity achieved with $\tau=1000$ ns is worse than the one obtained with $\tau=1100$ ns in Table \ref{tab:state-prep-fid-linear}. On the other hand, the longer pulse results in at least similar or better fidelity for all register sizes. This pattern suggests that longer pulses provide the optimizer more flexibility to adapt the shape of the pulse to the interaction between qubits and hardware constraints and find parameter configurations to overcome them.

\begin{table}
\centering
\begin{tabular}{|c|c|c|c|c|}
\hline
\textbf{\# of qubits} & \textbf{fidelity ($\tau=1000$ ns), \%} & \textbf{fidelity ($\tau=1200$ ns), \%}  \\
\hline
2 & $>99.99$ & $>99.99$  \\
3 & 98.06 & 99.35  \\
4 & 95.21 & 96.28  \\
5 & 91.19 & 99.86  \\
6 & 88.39 & 99.83  \\
7 & 85.36 & 99.76  \\
\hline
\end{tabular}
\caption{Fidelities of the state $\left|\psi_{\rm{final}}\right\rangle $ after the pulse optimization with different pulse durations.}
\label{tab:state-prep-fid-diff-durations}
\end{table}

Since interaction between qubits is an intrinsic property of the neutral-atom quantum devices, we also investigate the effect of different register configurations and inter-qubit distances on the optimized state fidelity. The results are presented in Table \ref{tab:state-prep-fid-diff-layouts}. Here we set the pulse duration $\tau=1100$ ns and the number of qubits $N=6$. The results show that stronger interaction generally leads to lower fidelity. The case with lattice constant $r_{12}=6.5\,\mu\rm{m}$ and rectangular register layout produces the lowest fidelity since the interaction between qubits is the strongest. Comparing the triangular and rectangular layouts with $r_{12}=7\,\mu\rm{m}$ we can see that the former results in a state with lower fidelity. This can be explained by the fact that in the triangular layout each qubit has at most 6 nearest neighbors separated by the lattice constant, whereas in the rectangular lattice there are only 4 such neighbors resulting in weaker average per-qubit interaction for the same lattice constant. 

\begin{table}
\centering
\begin{tabular}{|c|c|c|c|c|}
\hline
\textbf{layout} & \textbf{distance, $\rm{\mu m}$} & \textbf{fidelity, \%}  \\
\hline
linear & 7 &  99.85  \\
rect (2x3) & 7  & 97.75  \\
rect (2x3) & 6.5 & 88.69  \\
triangle & 7 & 95.73  \\
\hline
\end{tabular}
\caption{Fidelities of the state $\left|\psi_{\rm{final}}\right\rangle $ after the pulse optimization with different register layouts.}
\label{tab:state-prep-fid-diff-layouts}
\end{table}

\section{Conclusions and perspectives}
\label{sec:conc}

This paper introduces \pd, a \p extension enabling differentiability through integration with \tor, to streamline and leverage standard ML techniques for pulse sequence optimization targeting neutral-atom QPUs. We applied it to several research use cases aimed at global rotation optimizations and state preparation as a proof-of-principle to demonstrate its versatility in optimizing pulses in various cases, ranging from unconstrained piecewise constant pulses to constrained arbitrary custom waveforms and state preparation. In all cases, \pd produces pulses that realize unitaries with remarkably high fidelities compared to ideal ones. Due to its flexibility and efficiency, \pd opens the way to a wealth of pulse optimization research for analog quantum computing. For instance, the generation of noise-resilient pulses for optimal execution on hardware. Moreover, many high-level applications around combinatorial optimization for Quadratic Unconstrained Binary Optimization \cite{qubo2022} or Maximum Independent Set \cite{ebadi2022, wurtz2024, pasqalmis} problems or quantum graph machine learning around the Quantum Evolution Kernel \cite{albrecht2023} (see \cite{pasqalqek} for an open-source implementation for Pasqal's hardware) designed to be executed on Pasqal's QPU require optimal pulse shaping where \pd is expected to play a major role. Last but not least, we plan to address the main \pd limitation in the number of qubits by bridging the gap to high-performance differentiable analog emulators based on Matrix-Product States \cite{pasqalemusv} to explore optimal register topologies at scale.\\

\textbf{Acknowledgements.} Pasqal's team acknowledges funding from the European Union the projects EQUALITY (Grant Agreement 101080142) and PASQuanS2.1 (HORIZON-CL4-2022-QUANTUM02-SGA, Grant Agreement 101113690).

\bibliographystyle{alpha}
\bibliography{sample}

\newpage

\numberwithin{equation}{section}
\appendix

\input{appendixA.tex}

\input{appendixB.tex}

\input{appendixC.tex}

\end{document}

%% file: appendixA.tex
\section{Implementation details} \label{appendix:appA}

Here we present the code snippets showing the implementation details of global state rotation and state preparation simulations.

\subsection{Preparation of the constant-pulse sequence}

\begin{shaded*}
\begin{lstlisting}[caption=Creation of the constant-pulse sequence for the global rotation optimization problem with \pd., label={code:const-pulse-seq-creation}, language=python]
# Define sequence parameters.
seq_duration = 1050
n_pulses = 8

# Create a sequence and declare channels.
seq = Sequence(reg, MockDevice)
seq.declare_channel("rydberg_global", "rydberg_global")

# Declare sequence variables to be optimized as a dict.
seq_vars = {}
seq_vars["amp_params"] = [seq.declare_variable(f"amp_param_{i}") for i in range(n_pulses)]
seq_vars["det_params"] = [seq.declare_variable(f"det_param_{i}") for i in range(n_pulses)]
seq_vars["phase_params"] = [seq.declare_variable(f"phase_param_{i}") for i in range(n_pulses)]

# Add parameterized constant pulses to the sequence.
for i in range(n_pulses):
    seq.add(
        Pulse.ConstantPulse(
            duration=seq_duration // n_pulses,
            amplitude=seq_vars["amp_params"][i], 
            detuning=seq_vars["det_params"][i], 
            phase=seq_vars["phase_params"][i]
        ), 
        "rydberg_global"
    )
\end{lstlisting}
\end{shaded*}

\subsection{Initialization of the \qm for the constant-pulse sequence}

\begin{shaded*}
\begin{lstlisting}[caption=Definition of the optimizable parameters and hardware constraints for the \qm instance., label={code:const-pulse-seq-creation}, language=python]
# Extract variable names for the optimizable parameters.
var_names = [var.var.name for var_list in seq_vars.values() for var in var_list]

# Create a constraints dict (phase is unconstrained since it's related to the detuning).
constraints = {}
for name in var_names:
    if "amp" in name:
        # Apply device amplitude constraint.
        constraints[name] = {"min": 0.0, "max": int(MockDevice.channels["rydberg_global"].max_amp)}
    if "det" in name:
        # Apply device detuning constraint.
        constraints[name] = {"min": -MockDevice.channels["rydberg_global"].max_abs_detuning, "max": MockDevice.channels["rydberg_global"].max_abs_detuning}

# Set initial values for the optimizable parameters. Values are fixed for better convergence.
trainable_params = {name: torch.tensor(5.0) for name in var_names}

# Create batch tensor of all possible initial states.
init_state = torch.eye(2 ** n_qubits)

model = QuantumModel(
    seq,
    sampling_rate=0.05,
    trainable_param_values=trainable_params,
    constraints=constraints,
    solver=SolverType.DP5_SE,
    initial_state=init_state)
\end{lstlisting}
\end{shaded*}

\subsection{Preparation of the custom-pulse sequence}

\begin{shaded*}
\begin{lstlisting}[caption=Creation of the custom-pulse sequence for the global rotation optimization problem with \pd., label={code:cust-pulse-seq-creation}, language=python]
# Define sequence parameters.
duration = 1100
n_param = 20

# Create a sequence and declare channels.
seq = Sequence(reg, MockDevice)
seq.declare_channel("rydberg_global", "rydberg_global")

# Define a custom-shaped pulse.
amp_custom_param = seq.declare_variable("amp_custom", size=duration)
det_custom_param = seq.declare_variable("det_custom", size=duration)
cust_amp = CustomWaveform(amp_custom_param)
cust_det = CustomWaveform(det_custom_param)
pulse_custom = Pulse(cust_amp, cust_det, 0.0)

# Add pulse to the sequence.
seq.add(pulse_custom, "rydberg_global")
\end{lstlisting}
\end{shaded*}

\subsection{Initialization of the \qm for the custom-pulse sequence}

\begin{shaded*}
\begin{lstlisting}[caption=Definition of the optimizable parameters and hardware constraints for the \qm instance., label={code:const-pulse-seq-creation}, language=python]
# Define custom waveform parameters.
gamma = 0.05

# Create the sine interpolation matrix.
interp_mat = interpolate_sine(n_param, duration)

def custom_wf_amp(params):
    return torch.matmul(interp_mat, int(MockDevice.channels["rydberg_global"].max_amp) * torch.sigmoid(gamma * params))

def custom_wf_det(params):
    return torch.matmul(interp_mat, int(MockDevice.channels["rydberg_global"].max_abs_detuning) * torch.tanh(gamma * params))

# Define pulse parameters.
amp_values = 5 * torch.rand(n_param, requires_grad=True) - 2.5
det_values = 5 * torch.rand(n_param, requires_grad=True) - 2.5
trainable_params = {
    "amp_custom": ((amp_values,), custom_wf_amp),
    "det_custom": ((det_values,), custom_wf_det),
}

# Create a quantum model from the sequence.
model = QuantumModel(seq, trainable_params, sampling_rate=0.05, solver=SolverType.DP5_SE, initial_state=init_state)
\end{lstlisting}
\end{shaded*}

\newpage

\subsection{Optimization loop for the global rotation and state preparation problems}

\begin{shaded*}
\begin{lstlisting}[caption=Optimization loop used for global rotation and state preparation simulations., label={code:opt-loop-glob-rot-state-prep}, language=python]
# Initialize the optimizer and the learning rate scheduler.
optimizer = torch.optim.Adam(model.parameters(), lr=5.0)
scheduler = torch.optim.lr_scheduler.CosineAnnealingLR(optimizer, T_max=50)
epochs = 1000
min_change = 0.01
num_loss_plateu = 6

loss_dict = {}
for t in range(epochs):
    # Calculate the loss with the final state.
    _, final_state = model.forward()
    loss = 1 - fidelity(target_state, final_state[-1])
    
    # Perform backpropagation.
    loss.backward()
    optimizer.step()
    optimizer.zero_grad()

    # Log the loss value together with model params.
    loss_dict[t] = {"loss": float(loss), "params": {name: param.data.clone().detach() for name, param in model.named_parameters()}}

    if len(loss_dict) > num_loss_plateu and loss > 0.1:
        last_losses = [loss_dict[i]["loss"] for i in range(t-num_loss_plateu, t + 1)]
        diffs = [abs(last_losses[i] - last_losses[i - 1]) for i in range(-1, -num_loss_plateu-1, -1)]
        if all(diff < min_change for diff in diffs):
            # Update the learning rate.
            scheduler.step(epoch=0)
        else:
            # Update the learning rate.
            scheduler.step()
    else:
        # Update the learning rate.
        scheduler.step()

    if loss < 0.0001:
        break

    # Update the sequence with changed pulse parameter values.
    model.update_sequence()

# Get the best parameter set.
sorted_losses = dict(sorted(loss_dict.items(), key=lambda x: x[1]["loss"]))
best_param_set = list(sorted_losses.values())[0]["params"]
\end{lstlisting}
\end{shaded*}

%% file: appendixB.tex
\section{Optimized sequences for the gate optimization problem} \label{appendix:appB}

Here we present the visualizations of the optimized sequences for the gate optimization problem.

\subsection{Constant-pulse sequences}

\begin{figure}
\centering
\includegraphics[width=\textwidth]{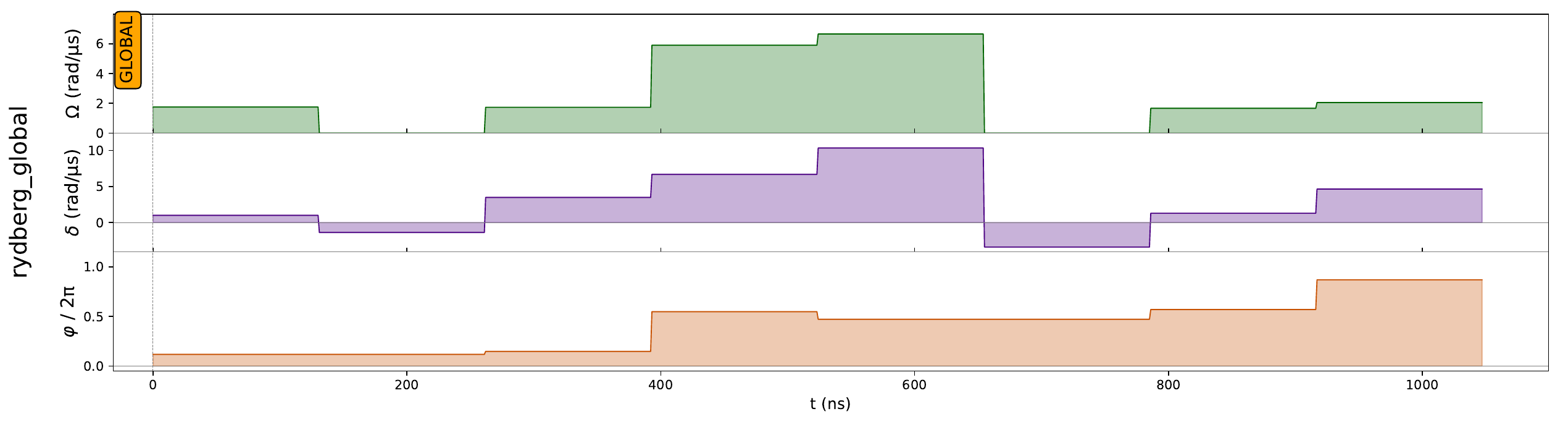}
\caption{Visualization of the optimized sequence consisting of the constant-waveform pulse for the gate optimization problem with a linear 3-qubit register.}
\label{fig:3q-gate-opt-const-app}
\end{figure}

\begin{figure}
\centering
\includegraphics[width=\textwidth]{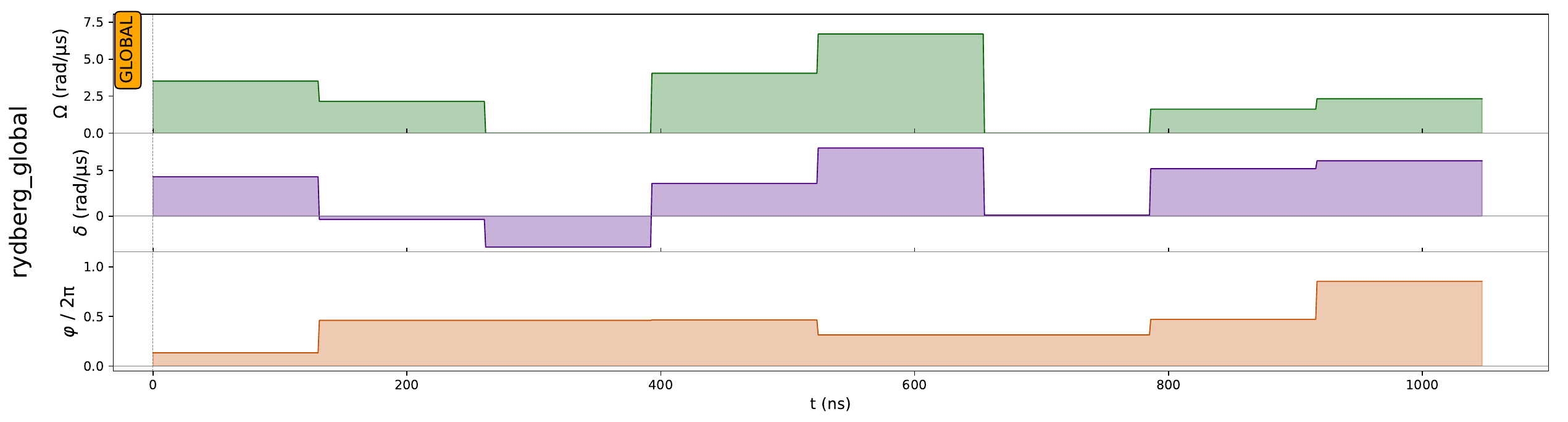}
\caption{Visualization of the optimized sequence consisting of the constant-waveform pulse for the gate optimization problem with a linear 4-qubit register.}
\label{fig:4q-gate-opt-const-app}
\end{figure}

\begin{figure}
\centering
\includegraphics[width=\textwidth]{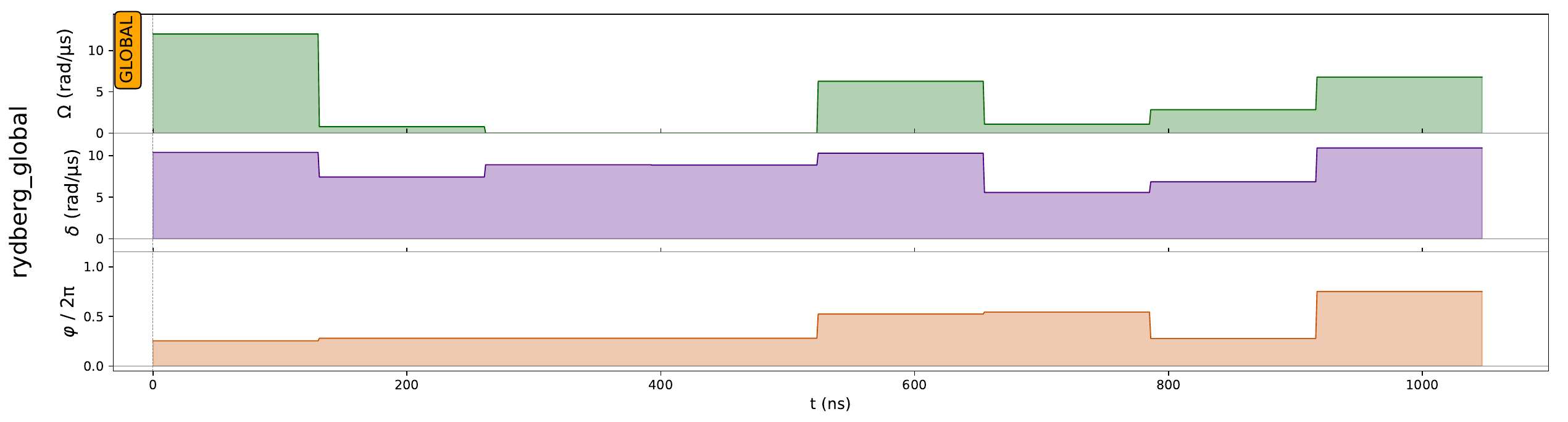}
\caption{Visualization of the optimized sequence consisting of the constant-waveform pulse for the gate optimization problem with a linear 5-qubit register.}
\label{fig:5q-gate-opt-const-app}
\end{figure}

\begin{figure}
\centering
\includegraphics[width=\textwidth]{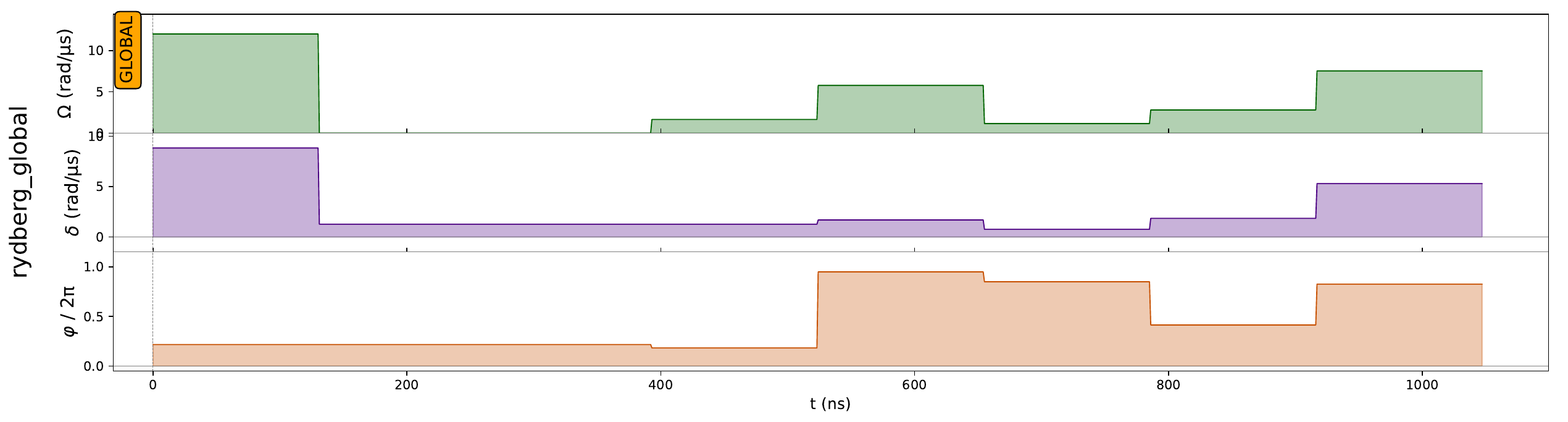}
\caption{Visualization of the optimized sequence consisting of the constant-waveform pulse for the gate optimization problem with a linear 6-qubit register.}
\label{fig:6q-gate-opt-const-app}
\end{figure}

\begin{figure}
\centering
\includegraphics[width=\textwidth]{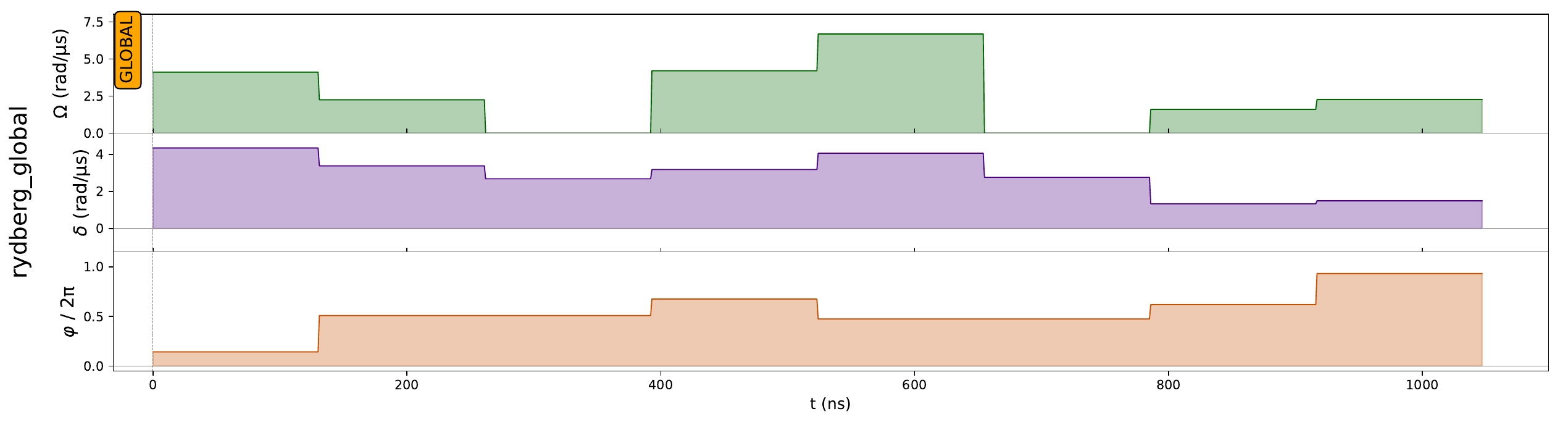}
\caption{Visualization of the optimized sequence consisting of the constant-waveform pulse for the gate optimization problem with a linear 7-qubit register.}
\label{fig:7q-gate-opt-const-app}
\end{figure}

\subsection{Custom-pulse sequences}

\begin{figure}
\centering
\includegraphics[width=\textwidth]{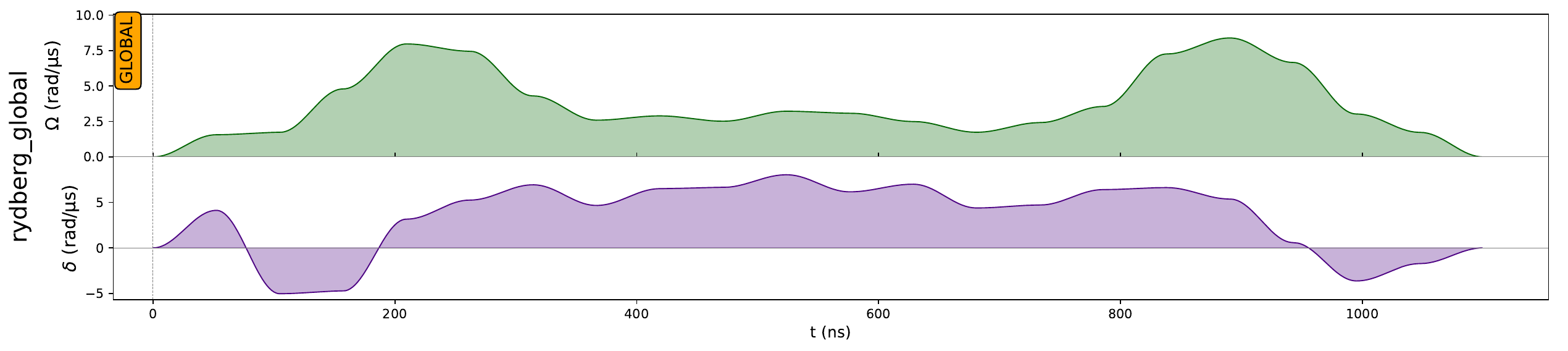}
\caption{Visualization of the optimized sequence consisting of the custom-waveform pulse for the gate optimization problem with a linear 2-qubit register.}
\label{fig:2q-gate-opt-cust-app}
\end{figure}

\begin{figure}
\centering
\includegraphics[width=\textwidth]{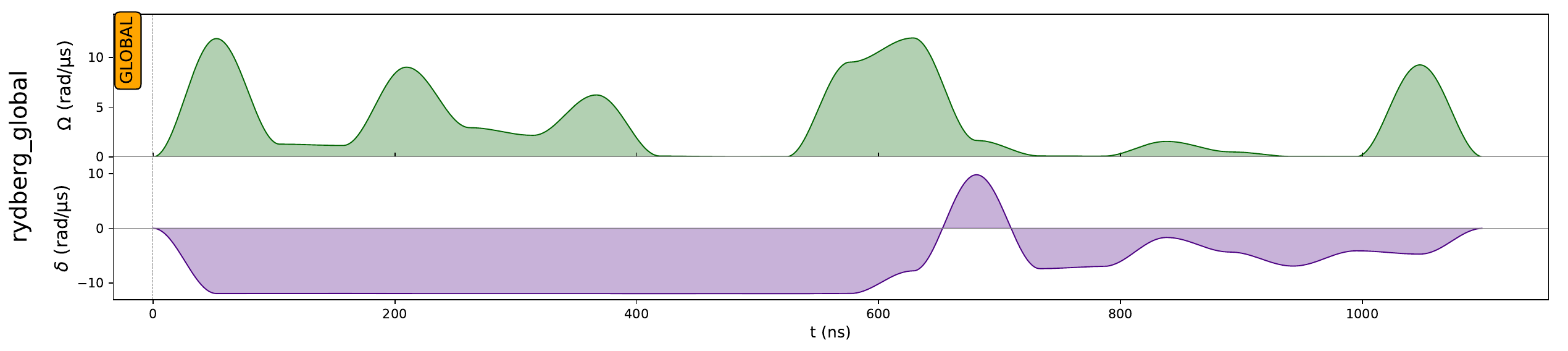}
\caption{Visualization of the optimized sequence consisting of the custom-waveform pulse for the gate optimization problem with a linear 3-qubit register.}
\label{fig:3q-gate-opt-cust-app}
\end{figure}

\begin{figure}
\centering
\includegraphics[width=\textwidth]{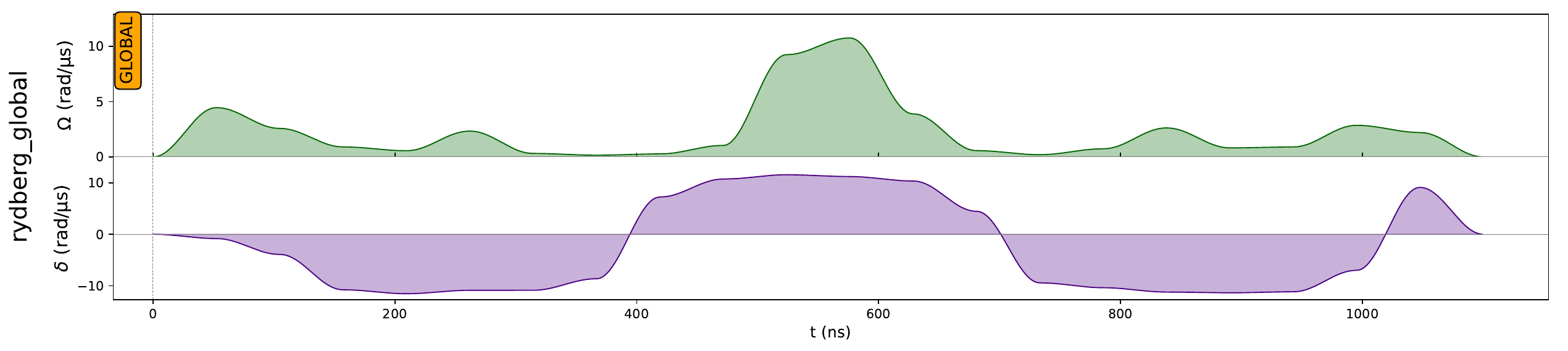}
\caption{Visualization of the optimized sequence consisting of the custom-waveform pulse for the gate optimization problem with a linear 5-qubit register.}
\label{fig:5q-gate-opt-cust-app}
\end{figure}

\begin{figure}
\centering
\includegraphics[width=\textwidth]{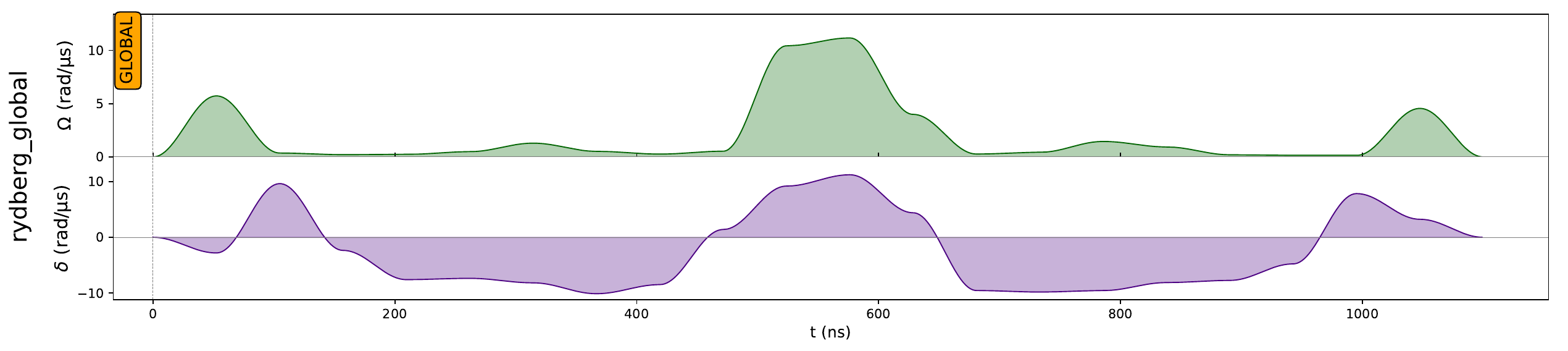}
\caption{Visualization of the optimized sequence consisting of the custom-waveform pulse for the gate optimization problem with a linear 6-qubit register.}
\label{fig:6q-gate-opt-cust-app}
\end{figure}

\begin{figure}
\centering
\includegraphics[width=\textwidth]{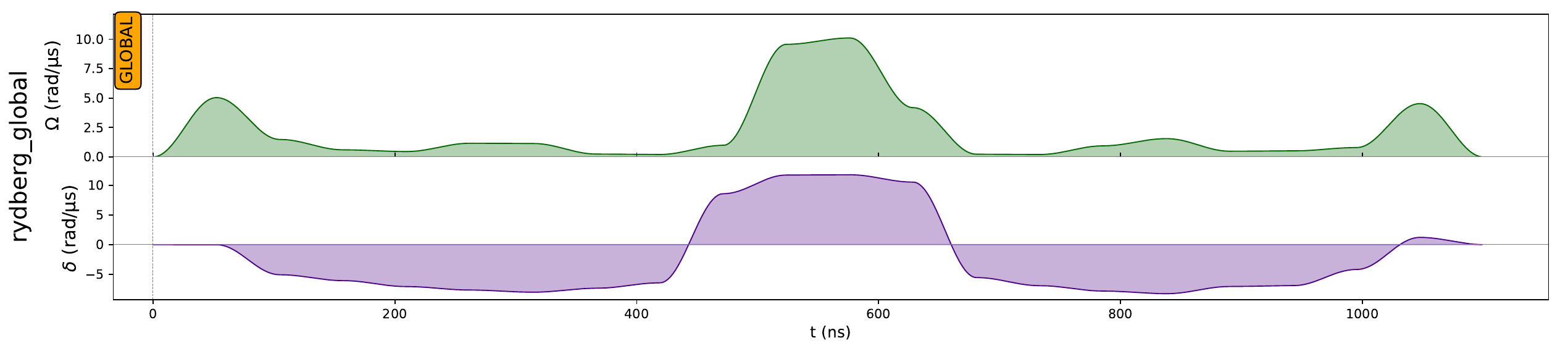}
\caption{Visualization of the optimized sequence consisting of the custom-waveform pulse for the gate optimization problem with a linear 7-qubit register.}
\label{fig:7q-gate-opt-cust-app}
\end{figure}

%% file: appendixC.tex
\section{Optimized sequences for the state preparation problem} \label{appendix:appC}

Here we present the visualizations of the optimized sequences for the state preparation problem.

\subsection{Sequences with different register size}

\begin{figure}
\centering
\includegraphics[width=\textwidth]{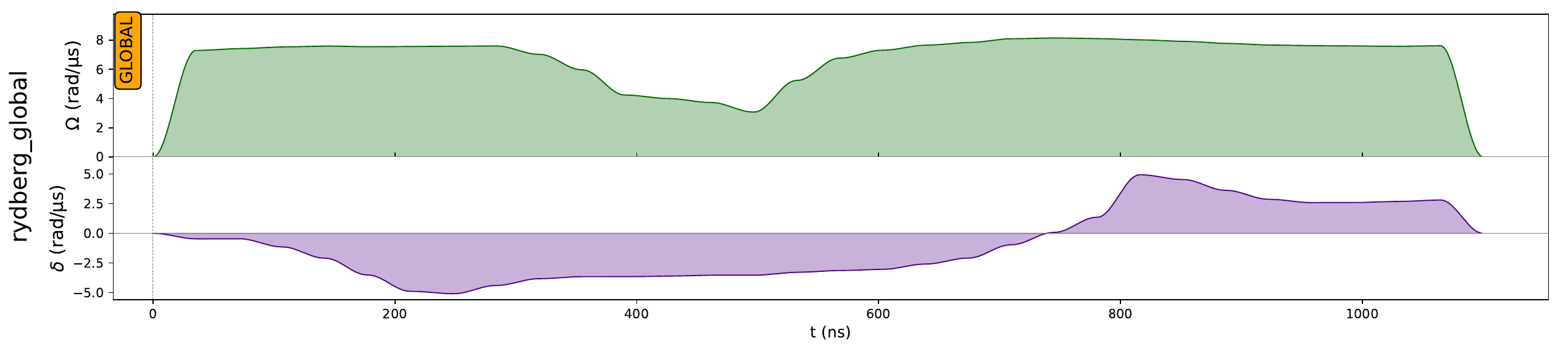}
\caption{Visualization of the optimized sequence consisting of the custom-waveform pulse for the state preparation problem with a linear 2-qubit register.}
\label{fig:2q-state-prep-cust-app}
\end{figure}

\begin{figure}
\centering
\includegraphics[width=\textwidth]{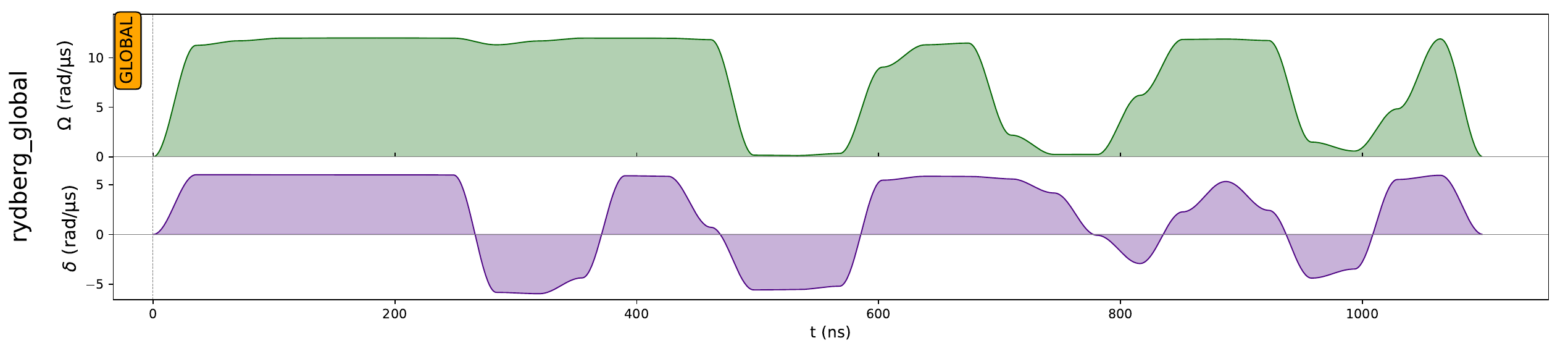}
\caption{Visualization of the optimized sequence consisting of the custom-waveform pulse for the state preparation problem with a linear 3-qubit register.}
\label{fig:3q-state-prep-cust-app}
\end{figure}

\begin{figure}
\centering
\includegraphics[width=\textwidth]{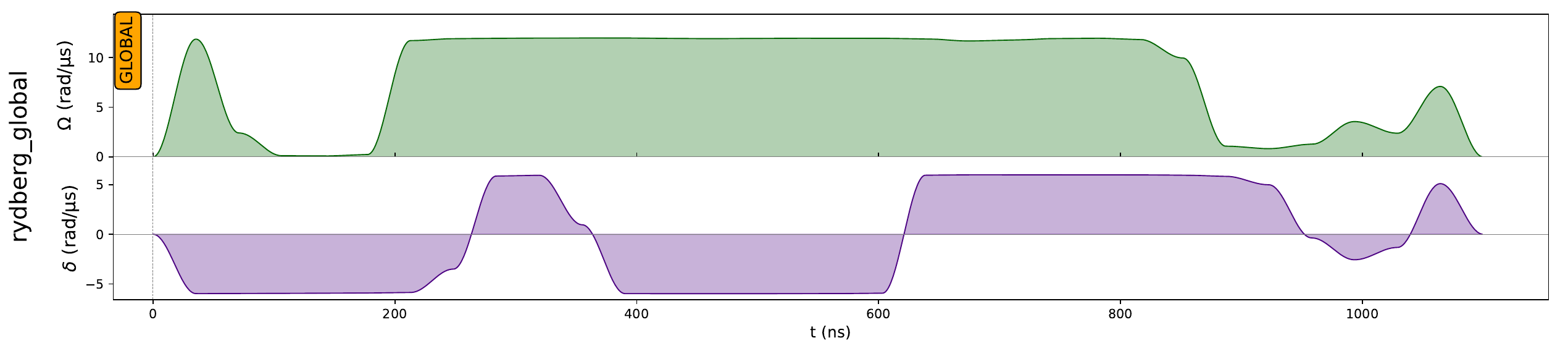}
\caption{Visualization of the optimized sequence consisting of the custom-waveform pulse for the state preparation problem with a linear 4-qubit register.}
\label{fig:4q-state-prep-cust-app}
\end{figure}

\begin{figure}
\centering
\includegraphics[width=\textwidth]{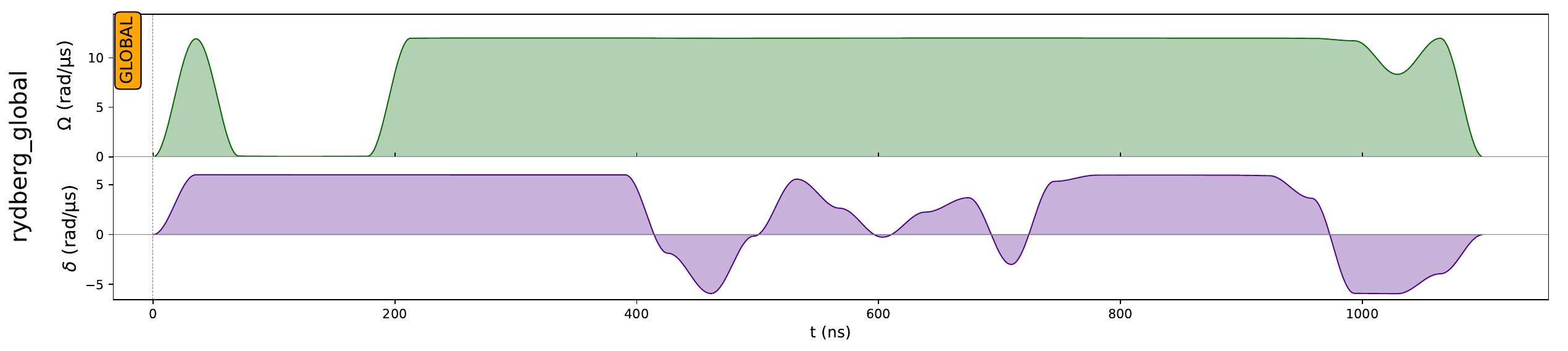}
\caption{Visualization of the optimized sequence consisting of the custom-waveform pulse for the state preparation problem with a linear 5-qubit register.}
\label{fig:5q-state-prep-cust-app}
\end{figure}

\begin{figure}
\centering
\includegraphics[width=\textwidth]{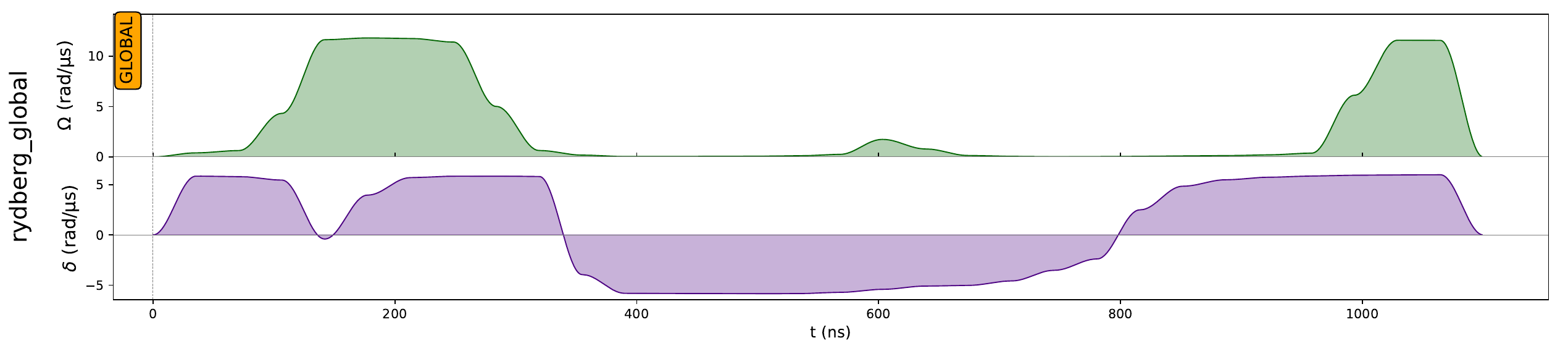}
\caption{Visualization of the optimized sequence consisting of the custom-waveform pulse for the state preparation problem with a linear 7-qubit register.}
\label{fig:7q-state-prep-cust-app}
\end{figure}

\subsection{Sequences with different register layouts}

\begin{figure}
\centering
\includegraphics[width=\textwidth]{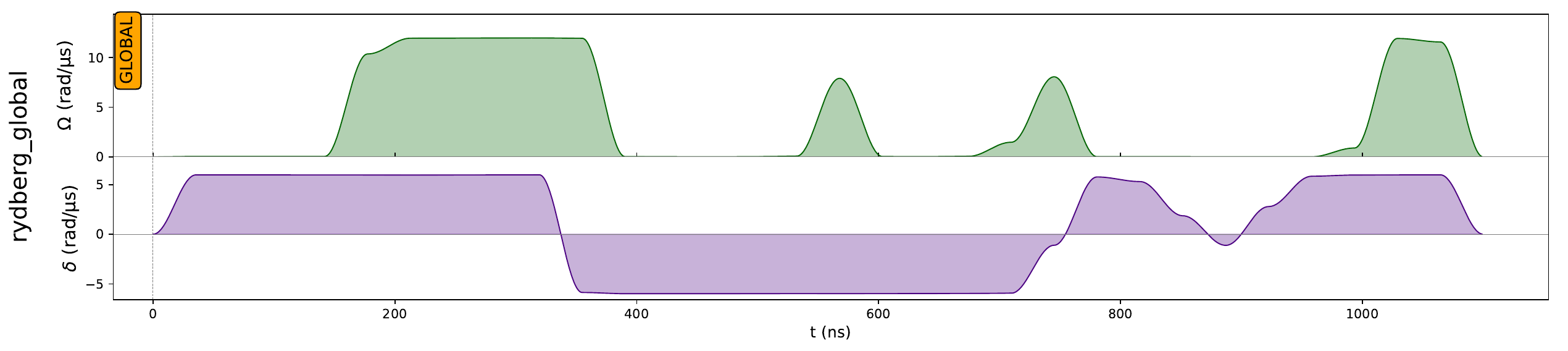}
\caption{Visualization of the optimized sequence consisting of the custom-waveform pulse for the state preparation problem with a rectangular (2x3) 6-qubit register and inter-qubit distance $r_{12}=7\,\mu\rm{m}$.}
\label{fig:4q-state-prep-cust-app}
\end{figure}

\begin{figure}
\centering
\includegraphics[width=\textwidth]{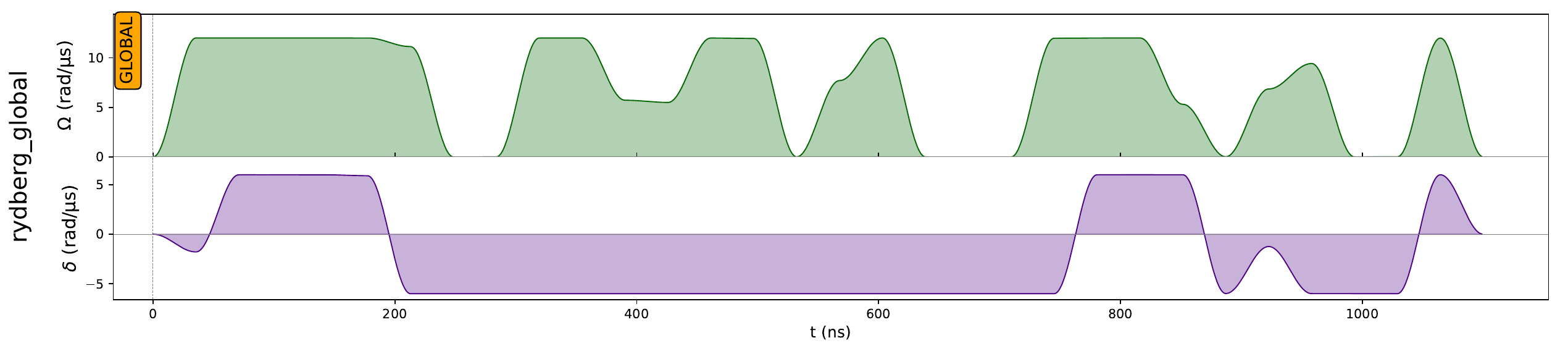}
\caption{Visualization of the optimized sequence consisting of the custom-waveform pulse for the state preparation problem with a rectangular (2x3) 6-qubit register and inter-qubit distance $r_{12}=6.5\,\mu\rm{m}$.}
\label{fig:5q-state-prep-cust-app}
\end{figure}

\begin{figure}
\centering
\includegraphics[width=\textwidth]{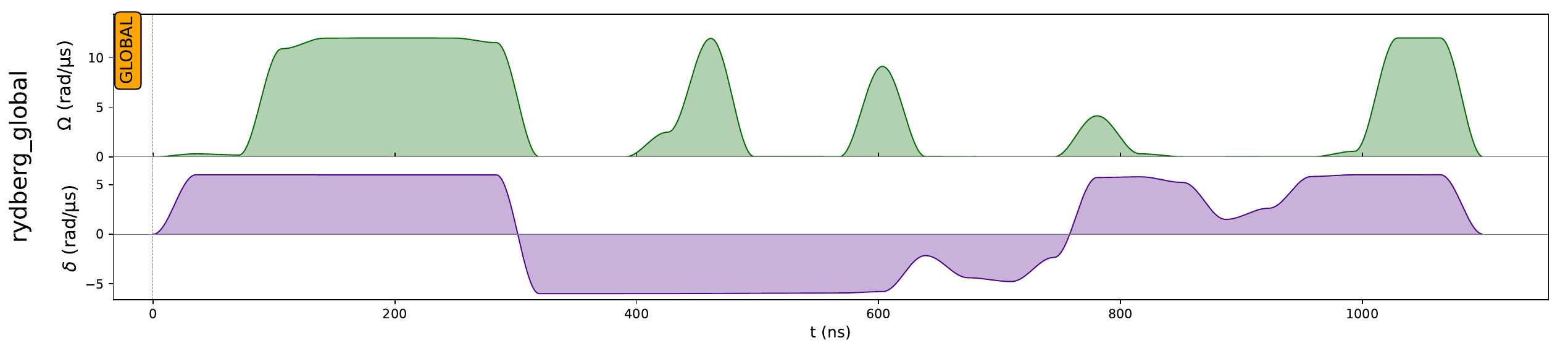}
\caption{Visualization of the optimized sequence consisting of the custom-waveform pulse for the state preparation problem with a triangular 6-qubit register and inter-qubit distance $r_{12}=7\,\mu\rm{m}$.}
\label{fig:7q-state-prep-cust-app}
\end{figure}